\documentclass[aps,prd,preprint,superscriptaddress]{revtex4}
\usepackage{float,graphicx}

\begin{document}


\title{Dark Energy as a Signature of Extra Dimensions}



\author{B.~Li }
\email[Email address: ]{bli@phy.cuhk.edu.hk}
\affiliation{Department of Physics, The Chinese University of Hong
Kong, Hong Kong SAR, China}

\author{M.~-C.~Chu}
\email[Email address: ]{mcchu@phy.cuhk.edu.hk}
\affiliation{Department of Physics, The Chinese University of Hong
Kong, Hong Kong SAR, China}

\author{K.~C.~Cheung}
\affiliation{Department of Physics, The Chinese University of Hong
Kong, Hong Kong SAR, China}

\author{A.~Tang}
\affiliation{Department of Physics, The Chinese University of Hong
Kong, Hong Kong SAR, China}


\date{\today}

\begin{abstract}
One of the most important and surprising discoveries in cosmology
in recent years is the realization that our Universe is dominated
by a mysterious dark energy, which leads to an accelerating
expansion of space-time. A simple generalization of the standard
Friedmann-Robertson-Walker equations based on General Relativity
and the Cosmological Principles with the inclusion of a number of
closed extra dimensions reproduces the currently observed data on
dark energy, without the introduction of any cosmological constant
or new particles. In particular, with a few extra dimensions, we
obtain the redshift dependence of the deceleration parameter, the
dark energy equation of state, as well as the age of the Universe
in agreement with data, with essentially no free parameter. This
model predicts that the extra dimensions have been compactified
throughout the cosmic history, and it therefore suggests that
signals from early Universe may give promising signatures of extra
dimensions.
\end{abstract}


\maketitle


Recent advances in cosmological observations produce a large set
of high quality data from which an unprecedented detailed
knowledge of the universe can be extracted.  Furthermore, a
$'$cosmic concordance$'$\cite{Tegmark2001} has emerged from
several different and independent observations such as Type IA
supernovae \cite{Riess1998, Perlmutter1999}, cosmic microwave
background (CMB) anisotropies \cite{Spergel2003} plus large scale
galaxy surveys \cite {Scranton2003} and Sachs-Wolfe effects
\cite{Colless2001}, pointing to a universe with flat space-time
dominated by a mysterious dark energy, which accounts for about
70\% of the total energy content.  The remaining 30\% contribution
is largely due to dark matter.  Furthermore, the dark energy is
characterized by an equation of state (EOS),
$P_{\Lambda}=w\rho_{\Lambda}$, relating its pressure $P_{\Lambda}$
and energy density $\rho_{\Lambda}$, and $w<-1$ is concluded from
data \cite{DiPietro2003, Riess2004, Alcaniz2004}, which gives rise
to a repulsive force and thus an accelerating expansion of
space-time. The evolution of $w$, $w(z)$, has also been traced
\cite{DiPietro2003, Riess2004}, leading to the conclusion that the
universe had undergone an earlier deceleration before changing to
the current accelerating phase. In addition to the usual
$\Lambda$CDM model, many alternatives have been constructed to
account for this dark energy. These include various quintessence
models \cite{Wetterich1988, Peebles1988, Caldwell1998}, Chaplygin
gas model \cite{Kamenshchik2001}, modified gravity and
scalar-tensor theories \cite{Boisseau2000, Damour2002} and so on.
The nature of dark energy has become one of the most important
research topics in cosmology and contemporary physics.

On the other hand, theories of extra dimensions have received a
lot of attention in recent years perhaps because string theory
also requires more than four dimensions of space-time
\cite{Polchinski1998}. Various Braneworld models \cite{Csaki2004}
have been proposed, in which the extra dimensions need not be
small. The hope that these large extra dimensions may provide a
simple solution to the hierarchy problem and that they may be
observed by upcoming experiments has created much excitement in
this subject \cite{Hewett2002}. Cosmological models involving
extra dimensions \cite{Daffayet2002, Sahni2003, Gu2002, Gu2003,
Gu2004} to explain the current cosmic acceleration are also
beginning to appear.

In this article, we adopt the simplest generalization of general
relativistic cosmology by incorporating a few extra spatial
dimensions to show that such a minimal extension is already
adequate to reproduce the most important cosmological features,
\emph{i.e.} the observed age of the universe, $t_{0}$, and the
evolution of both $w$ and the deceleration parameter $q \equiv
-(\ddot{a}/a)(\dot{a}/a)^{-2}$. Based on these simple assumptions,
dark energy is shown to be a possible manifestation of extra
dimensions. It is further argued that the best chance to probe the
extra dimensions lies in the early universe \cite{Li2005}.

The base of the present model is a $1+3+n$ dimensional space-time
with $n$ being the number of extra dimensions. With the assumption
of homogeneity and isotropy in both the ordinary and the extra
dimensions, the generalized Robertson-Walker metric that describes
this space-time takes the following form:
\begin{equation}
ds^{2}=dt^{2}-a^{2}(t)\left(\frac{dr_{a}^{2}}{1-k_{a}r_{a}^{2}}+r_{a}^{2}d\Omega_{a}^{2}\right)
-b^{2}(t)\left(\frac{dr_{b}^{2}}{1-k_{b}r_{b}^{2}}+r_{b}^{2}d\Omega_{b}^{2}\right),
\end{equation}
where $r_{a} (r_{b}), \Omega_{a} (\Omega_{b})$ are the radial and
angular coordinates of ordinary (extra) dimensions, $a(t)$, $b(t)$
and $k_{a}, k_{b}$ are the scale factors and curvatures of the
ordinary three-dimensional space and the extra dimensions,
respectively. The matter content in the Universe is assumed to be
perfect fluid and the corresponding stress-energy tensor is given
by
\begin{equation}
T_{N}^{M}=diag \left(\bar{\rho}, -\bar{P_{a}}, -\bar{P_{a}},
-\bar{P_{a}}, -\bar{P_{b}}, -\bar{P_{b}},\cdot\cdot\cdot\right),
\end{equation}
where $\bar{\rho}$ denotes the high dimensional energy density and
$\bar{P_{a}}, \bar{P_{b}}$ the pressure in the ordinary and extra
dimensions. Then the Einstein equations lead to the following
$1+3+n$ dimensional Friedmann-Robertson-Walker (FRW) equations
($\bar{G}$ is the higher dimensional gravitational constant):
\begin{equation}
3\left[\left(\frac{\dot{a}}{a}\right)^{2}+\frac{k_{a}}{a^{2}}\right]=8\pi
\bar{G}\bar{\rho}+\tilde{\rho}_{eff},
\end{equation}
\begin{equation}
2\frac{\ddot{a}}{a}+\left[\left(\frac{\dot{a}}{a}\right)^{2}+\frac{k_{a}}{a^{2}}\right]=-8\pi
\bar{G}\bar{P_{a}}-\tilde{P_{a,}}_{eff},
\end{equation}
\begin{equation}
3\frac{\ddot{a}}{a}+3\left[\left(\frac{\dot{a}}{a}\right)^{2}+\frac{k_{a}}{a^{2}}\right]=-8\pi
\bar{G}\bar{P_{b}}-\tilde{P_{b,}}_{eff},
\end{equation}
in which
\begin{equation}
\tilde{\rho}_{eff}\equiv
-\frac{n\left(n-1\right)}{2}\left[\left(\frac{\dot{b}}{b}\right)^{2}+\frac{k_{b}}{b^{2}}\right]-3n\frac{\dot{a}}{a}\frac{\dot{b}}{b},
\end{equation}
\begin{equation}
\tilde{P_{a,}}_{eff}\equiv
n\frac{\ddot{b}}{b}+\frac{n\left(n-1\right)}{2}\left[\left(\frac{\dot{b}}{b}\right)^{2}+\frac{k_{b}}{b^{2}}\right]
+2n\frac{\dot{a}}{a}\frac{\dot{b}}{b},
\end{equation}
\begin{equation}
\tilde{P_{b,}}_{eff}\equiv
\left(n-1\right)\frac{\ddot{b}}{b}+\frac{\left(n-1\right)\left(n-2\right)}{2}\left[\left(\frac{\dot{b}}{b}\right)^{2}
+\frac{k_{b}}{b^{2}}\right]+3\left(n-1\right)\frac{\dot{a}}{a}\frac{\dot{b}}{b}.
\end{equation}
Recall that in theories involving extra dimensions the $1+3$ D
energy density and Newtonian gravitational constant $\rho,\ G_{N}$
are related with $\bar{\rho},\ \bar{G}$ by $\rho=\bar{\rho}V^{n}$
and $G_{N}=\bar{G}/V^{n}$ with $V^{n}$ being the volume of extra
dimensions \cite{ADD1999}, we could replace the
$\bar{G}\bar{\rho}$ in Eq. (3) by $G_{N}\rho$.

Obviously, the left-hand sides of Eqs. (3) and (4) are just the
same as those in the $1+3$ D FRW equations.  The effects of the
extra dimensions are summarized in $\tilde{\rho}_{eff}$ and
$\tilde{P_{a,}}_{eff}$, which could be interpreted as the energy
density and pressure of the geometry-induced matter first proposed
by Einstein \emph{et al.} \cite{Einstein1956, Salam1980} and later
developed by Wesson \emph{et al.} \cite{Overduin1997}. In this
sense the $1+3$ D FRW equations are reproduced as parts of the
higher dimensional ones and naturally we ask ourselves whether the
observed dark energy can be explained by the effects of the extra
dimensions.

The generalized FRW equations determine uniquely the evolutions of
$a$ and $b$ once a complete set of initial conditions is given. A
complete study of the various possibilities in the generalized FRW
model will be presented in Ref. 24.  The present work focuses on
the case of $k_{a}=0$, $k_{b}=1$, $\bar{P_{a}}=\bar{P_{b}}=0$ and
$n=7$. It is different from the model discussed in Ref. 23 where
flat extra dimensions and non-flat ordinary dimensions are
considered; the main results are also considerably different. The
choices of $k_{a}$ and $\bar{P_{a}}$ follow from observational
data, which imply that currently the ordinary dimensions are flat
and matter is mostly non-relativistic.  Our solutions do not
depend on $n$ sensitively and the case of $n=7$ is simply chosen
for the sake of illustration.

The initial conditions of the generalized FRW equations in Eqs.
(3)-(5) are taken from the current observed values of the Hubble
parameter $H_{0}=\dot{a}_{0}/a_{0}$, the deceleration parameter
$q_{0}=-\ddot{a}_{0}/a_{0}H_{0}^{2}$, and the cosmological matter
density parameter $\Omega_{m0}=8\pi G_{N}\rho_{0}/3H_{0}^{2}$,
where a subscript $'0'$ indicates present day $(z=0)$ value. The
Hubble parameter has been measured with many different techniques.
For instance, the HST Key project \cite{Freedman2001} gives an
estimation of $h=0.72\pm0.03\pm0.07$ $(h\equiv H_{0}/100 Km s^{-1}
Mpc^{-1})$ based on the traditional distance ladder approach, Treu
and Koopmans \cite{Treu2002} obtain
$h=0.59_{-0.07}^{+0.12}\pm0.03$ using gravitational lens time
delay method, Reese \emph{et al.} \cite{Reese2002} combines the
Sunyaev-Zel$'$dovich and X-ray flux measurements of galaxy
clusters to find $h=0.6\pm0.04_{-0.18}^{+0.13}$, and Jimenez
\emph{et al.} \cite{Jimenez2003} use the constraints from the
absolute ages of Galactic stars and the observed position of the
first peak in the CMB angular power spectrum to obtain $h=0.69
\pm0.12$. We adopt the range $0.58<h<0.70$ suggested by these
measurements. The current deceleration factor is extracted from
the allowed values in Ref. 8: $q_{0}\simeq -1.3\sim-0.4$ (with
$-0.6\gtrsim q_{0}\gtrsim -1.1$ at the $1\sigma$ C. L.).  Finally,
we adopt the typical value1 of $\Omega_{m0}=0.3$. With the values
of $h, q_{0}, \Omega_{m0}$ and thus $\dot{a}_{0}/a_{0},
\ddot{a}_{0}/a_{0}$ fixed, Eqs. (3)-(8) are used as simple
algebraic equations to solve for the values of
$\dot{b}_{0}/b_{0}$, $\ddot{b}_{0}/b_{0}$ and
$k_{b}/b_{0}^{2}=1/b_{0}^{2}$. The initial conditions are then
completely specified such that the same group of equations can be
integrated backward in time to obtain $a(t)$, $b(t)$, $q(t)$,
$w(t)\equiv \tilde{P_{a,}}_{eff}(t)/\tilde{\rho}_{eff}(t)$ and the
cosmic age $t_{0}$.

\begin{figure}[]
\includegraphics[scale=1.3]{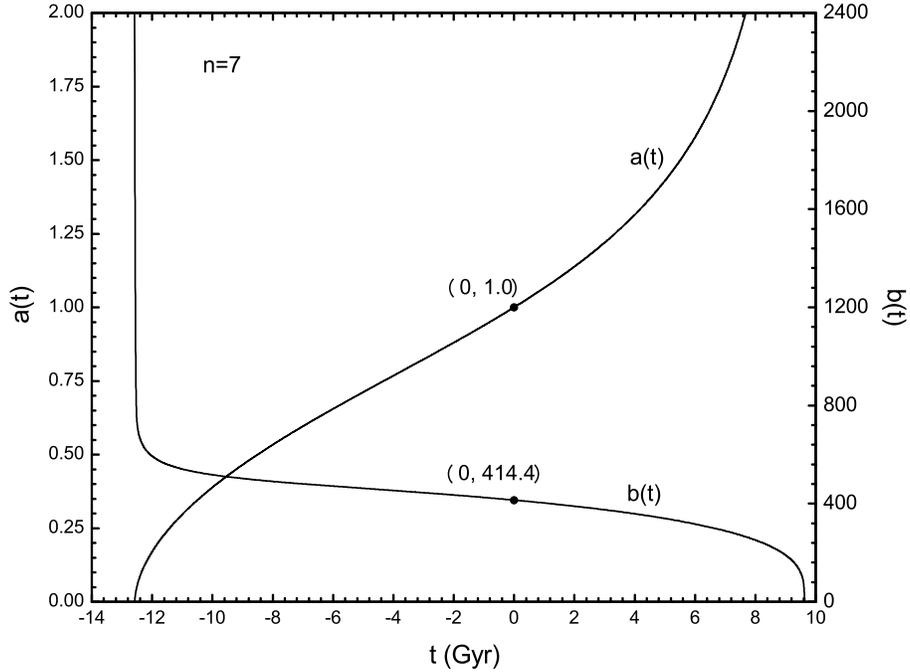}
\caption{The time evolution of $a$ and $b$ with $a_{0}$ set to be
1 by convention. Note that the largeness of $b$ is only the result
of normalizing the arbitrary (positive) $k_{b}$ to be 1 and such
that $a$ and $b$ are not comparable physically. The current time
is $t=0$.}
\end{figure}

\begin{figure}[]
\includegraphics[scale=1.3]{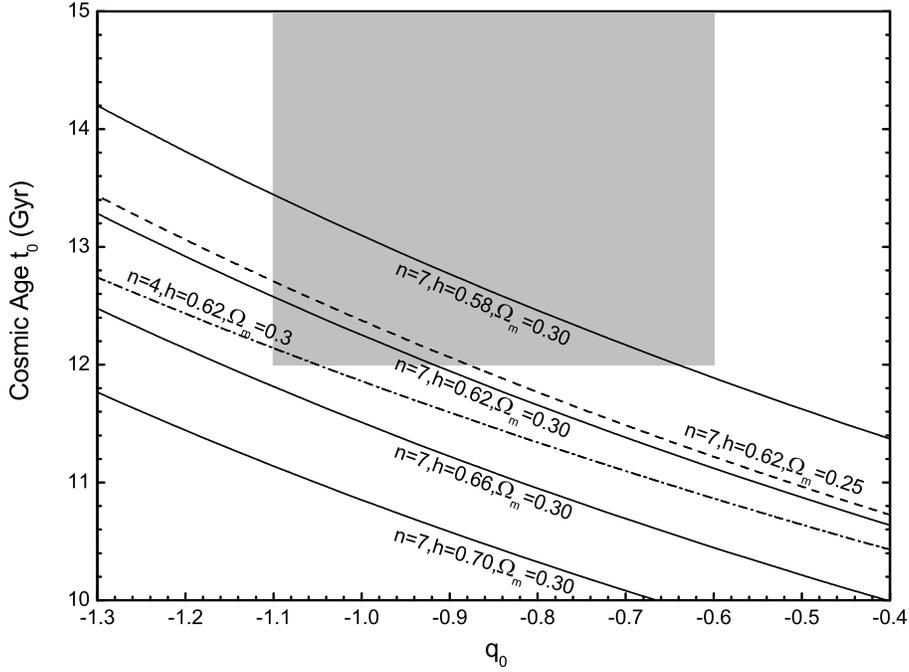}
\caption{The dependence of the model-predicted cosmic age $t_{0}$
on different choices of $q_{0}$, $h$, $\Omega_{m0}$ and $n$. The
grey patch represents the allowed area of the parameters according
to observations: $-0.6\gtrsim q_{0}\gtrsim-1.1$ as the $1\sigma$
C.~L. region given in Ref. 8 and $t_{0}$ is taken to be $\geq12.0\
Gyr$. The values of parameters are labelled aside.}
\end{figure}

Figure 1 shows the time evolutions of $a$ and $b$, with $h=0.62,\
q_{0}=-1.1$ and $a$ set to be 1 at present $(t = 0)$.
Qualitatively speaking, the present model reproduces the observed
deceleration-acceleration transition for the ordinary dimensions
in which the extra dimensions shrink continuously throughout the
cosmic history. The cosmic age $t_{0}$, \emph{i.e.} the time
between $a=0$ and $a=1$ (for our realistic purpose the duration of
the inflationary and radiation-dominated eras is so tiny compared
with the matter-dominated period that we could treat the latter as
$t_{0}$ approximately), is found from the figure to be $\simeq12.6
\ Gyr$ in good agreement with values quoted in literature: a lower
limit of cosmic age is given by the age of the oldest globular
clusters plus $0.2\sim0.3 \ Gyr$ \cite{Jimenez2003}, leading to a
typical value of $12 \ Gyr$ \cite{Feng2004}. To observe the
reality of our model, we show in Fig. 2 the dependence of
model-predicted $t_{0}$ on $q_{0}$ and $h$ for the $n=7$ case with
the observational bounds from $q_{0}$ and $t_{0}$ marked by grey
area. This result shows that a large allowed region exists that
gives solutions consistent with observed data. It is emphasized
that, although $n$ is a free chosen parameter, its variation does
not change the results significantly. For example, the cosmic age
in the case of $n=3$ differs from that for $n=7$ by only
$\sim5\%$.

\begin{figure}[]
\includegraphics[scale=1.3]{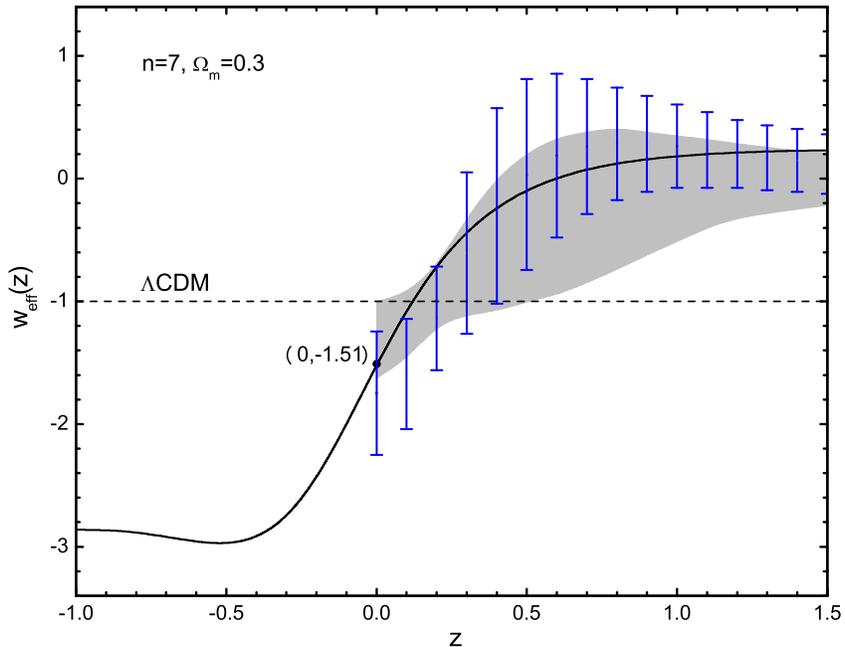}
\caption{The redshift dependence of dark energy EOS, $w(z)$. Two
sets of observational results $(z\geq0)$ extracted from Supernovae
data are represented by error bars and grey region
\cite{Alam2004a, Alam2004b} and the model prediction by solid
curve. Case of a cosmological constant $(w(z)=1)$ is denoted by
dashed line. Current state of the Universe in the $w-z$ plane is
also marked.}
\end{figure}

\begin{figure}[]
\includegraphics[scale=1.3]{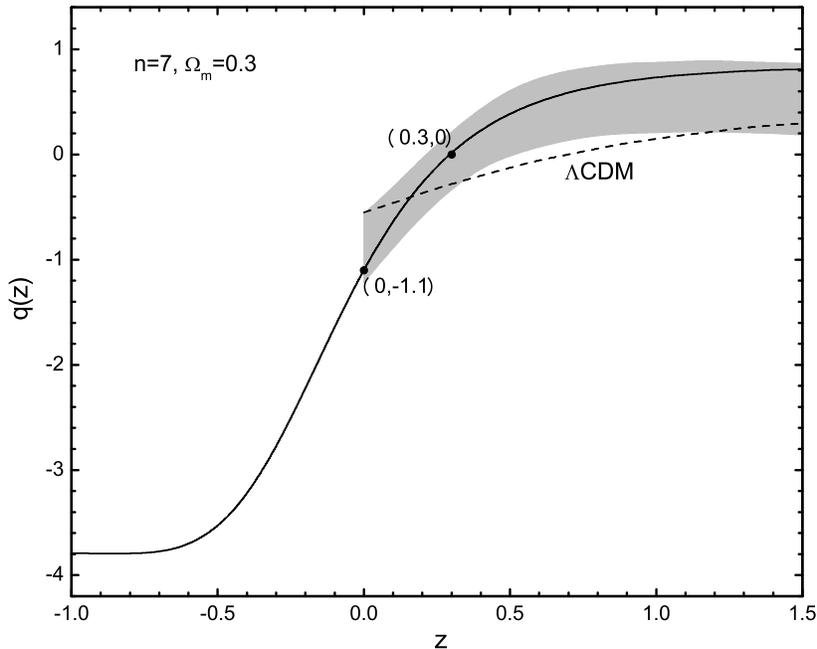}
\caption{Same as Fig. 3, but for the redshift dependence of $q$.
Grey region represents observational data \cite{Alam2004a,
Alam2004b} and solid (dashed) line the prediction of our
($\Lambda$CDM) model. The states of the Universe in the $q-z$
plane at current time and at the time of deceleration-acceleration
transition are marked.}
\end{figure}

\begin{figure}[]
\includegraphics[scale=0.9]{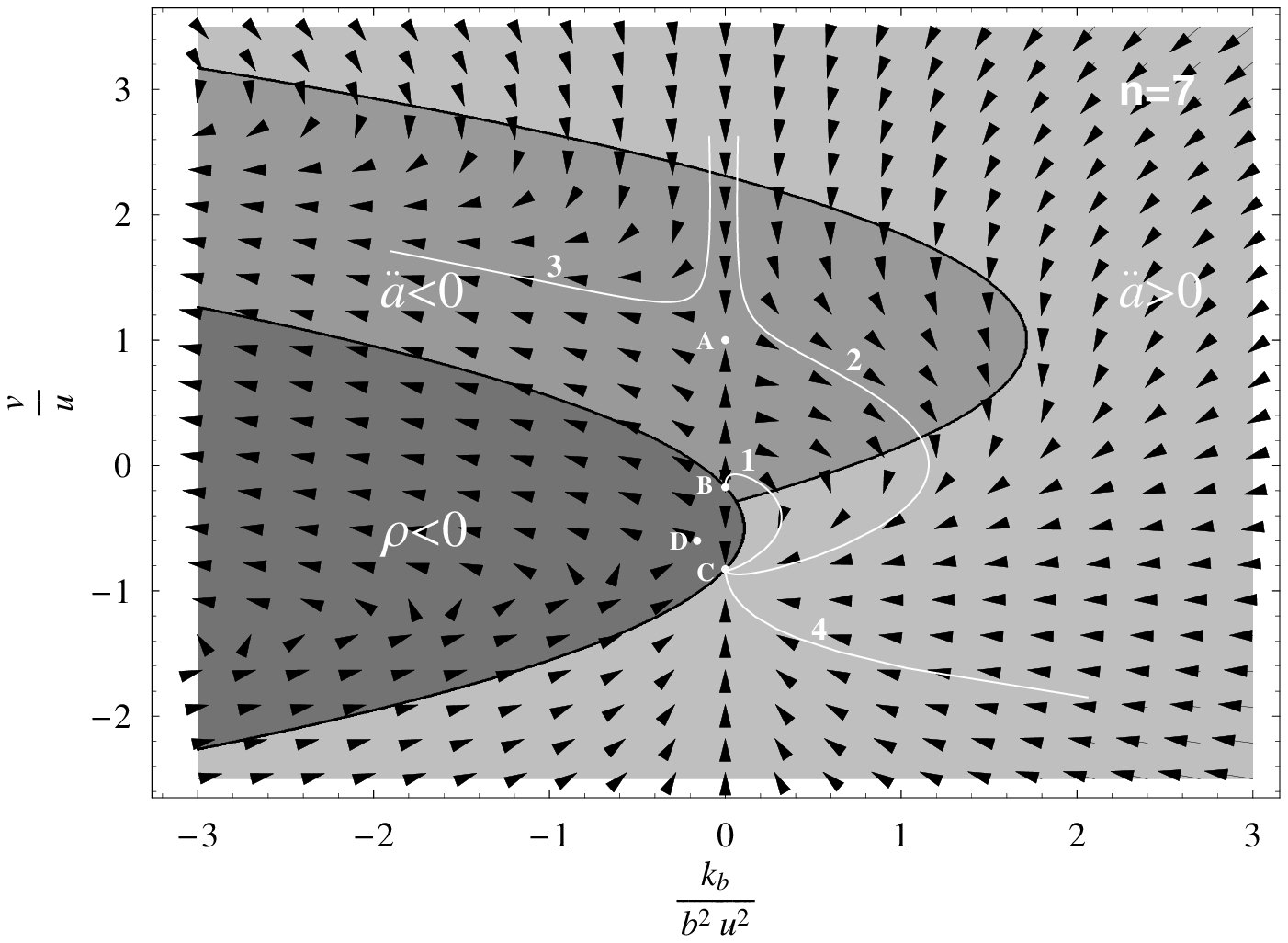}
\caption{Flow plot showing the evolution of solutions on the
$\left(X_{b}\equiv k_{b}/b^{2}u^{2},Y\equiv v/u\right)$ plane.
Regions corresponding to $\rho<(>)\ 0$ and $\ddot{a}<(>)\ 0$ are
shown. Four representative evolution paths are given by white
lines (1-4) and four fixed pointed are indicated (A-D).}
\end{figure}

A tighter constraint on a model of the dark energy is the redshift
dependence of its EOS, $w(z)$. For the chosen set of parameters
above, we show the $w(z)$ calculated from our model in Fig. 3. We
obtain $w_{0}=w(z=0)\simeq-1.5$ and $w'_{0}=dw(z=0)/dz\simeq4$.
These results agree with several recent analyses of supernova
data, including those given by De Pietro \& Claeskens
\cite{DiPietro2003}: $w_{0}\simeq-1.4, -12<w'_{0}<12$, Riess
\emph{et al.} \cite{Riess2004}: $w_{0}=-1.31_{-0.28}^{+0.22}$ and
other parametric reconstructions of $w(z)$ using the current
observational data \cite{Alam2004a, Alam2004b, Huterer2004,
Gong2004}, but not very well with the reported value of Ref. 8:
$w'_{0}=1.48_{-0.90}^{+0.81}$. However, the present results lie at
the boundary of the $1\sigma$ contour of the estimation using the
gold sample \emph{without} HST data in Ref. 8. This discrepancy
could be due to their simple assumption in the fitting relation
$w(z)=w_{0}+w'_{0}z$, which is valid only when $z$ is small.  If
$w(z)$ bends down as $z$ increases, as the present model (Fig. 3)
suggests, the said fitting will underestimate $w'_{0}$. This
underestimation may explain why the contours in the $w_{0}-w'_{0}$
plane shift downward when higher-redshift HST discovered data are
included in Ref. 8.  It is predicted that a lower $w'_{0}$ will be
found when even higher-redshift data are used.

Fig. 4 shows the redshift dependence of the deceleration parameter
$q(z)$.  The deceleration parameter is positive and nearly
constant $(\sim0.8)$ for large $z$ but changes rapidly to a
negative value $(\sim-3.8)$ between $z\sim1$ and $z\sim-0.5$ with
the transition from deceleration $(q>0)$ to acceleration $(q<0)$
occurring at $z\sim0.3$, consistent with the data-fitting result
given in Ref. 33.  The current state of the universe in the
$q_{0}-q'_{0}$ plane is around $(-1.1, 5)$, slightly outside the
$3\sigma$ C. L. according to Ref. 8, which again is expected to be
improved if the simple linear fitting is modified to include
higher order terms.

The robustness of the present model could be shown in the flow
plot \cite{Gu2004} of the solutions of Eqs. (3)-(8), as in Fig. 5
for a pressureless $n=7$ universe.  For $k_{a}=0$, the evolution
of any solution can be represented on a 2D plane formed by the two
parameters $\left(X_{b}\equiv k_{b}/b^{2}u^{2}, Y\equiv
v/u\right)$ where $u\equiv\dot{a}/a$ and $v\equiv\dot{b}/b$. The
arrows indicate the direction of time evolution. Four
representative paths are shown as white lines labelled 1-4, and
the regions corresponding to $\rho>(<) \ 0$ and $\ddot{a}>(<) \ 0$
are also marked. Only solutions flowing in the $\rho>0$ region are
considered. There are four finite fixed points, labelled A-D, with
C being the only stable one. Both of the flow patterns 1 and 2
possess a transition from deceleration to acceleration in
agreement with observational results while they differ in that
pattern 2 also exhibits an initial acceleration phase. However,
the current Hubble constant and deceleration parameter indicate
that pattern 1 offers a better description of the observable
universe, because the cosmic age associated with the pattern 2
solution is typically much shorter than the lower bound imposed by
globular cluster age \cite{Spergel2003, Feng2004}. It is noted
that the early deceleration to later acceleration evolution of
ordinary dimensions in our model is a robust feature with respect
to small perturbations in the cosmological parameters and, again,
varying $n$ does not alter the flow patterns significantly. But on
the other hand, the curvature of the extra dimensions is critical:
solutions corresponding to $k_{b}\leq0$ exhibit $(1)$ either a
strictly accelerating or decelerating phase or $(2)$ a transition
from acceleration to deceleration in contradiction to observation.

The concept of a running Newtonian gravitational constant has
received some attention in recent literature. If gravity
propagates freely in all of the ordinary and extra dimensions, the
effective 4D Newtonian constant $G_{N}$ should vary with time
according to \cite{ADD1999}
\begin{equation}
\frac{\dot{G_{N}}}{G_{N}}=-n\frac{\dot{b}}{b}.
\end{equation}
It seems that Eq. (9) poses a severe constraint \cite{Cline2003}
on $\dot{b}/b$. However, It is doubtable whether constraints on
$\dot{G}_{N}/G_{N}$ derived from laboratory, solar system, pulsar
tests and so on \cite{Uzan2003} are applicable to our cosmological
model: these tests are dominated by local gravitational fields and
are more appropriately described by static metrics, rather than
the cosmological metric given in Eq. (1). As a result, they do not
$'$feel$'$ the effect of cosmological changes such as $\dot{a}$.
And similarly, $\dot{b}$, being cosmological in nature, will be
not easily probed by these tests. So we will not consider this
specious constraint \cite{Comment1} on the time-evolution of extra
dimensions in our model.

The results of the present work may have important implications in
cosmology.  First of all, the present model shows that the extra
dimensions compactify throughout the cosmic history, with an
exponential deflationary phase extending up to $z\sim10$.  It
suggests that the most prominent signature of extra dimensions is
likely to be found in cosmological signals, such as CMB. Secondly,
the main point of this article is that dark energy is a possible
manifestation of extra dimensions.  Lastly, the fate of the
universe in our model is characterized by a continuously
accelerating expansion of the ordinary dimensions accompanied by
accelerating \emph{compactification} of the extra dimensions.
Eventually, $b$ becomes so small that quantum gravity dominates.

In summary, we have shown that a generalized FRW cosmology which
includes a few extra closed spatial dimensions predicts the
observed expansion history of the universe without any
cosmological constant. Dark energy arises as the effect of the
evolution of the extra dimensions, and its observed equation of
state is reproduced with essentially no free parameter.  The
solution is robust.  It suggests that much more prominent effects
of the extra dimensions may be observable in cosmic signals from
the early universe.

\begin{acknowledgments}
The work described in this paper was substantially supported by a
grant from the Research Grants Council of the Hong Kong Special
Administrative Region, China (Project No. 400803) and a Chinese
University Direct Grant (Project No. 2060248). A. Tang thanks a
Postdoctoral Fellowship from the Chinese University of Hong Kong.
\end{acknowledgments}

\newcommand{\noopsort}[1]{} \newcommand{\printfirst}[2]{#1}
  \newcommand{\singleletter}[1]{#1} \newcommand{\switchargs}[2]{#2#1}


\begin{thebibliography}{19}
\bibitem[{\citenamefont{Comment1}(2004)}]{Comment1}
  In our model with the quoted parameters, assuming the constraint from Eq. (9) is
  applicable and using a conservative observational
  result as $|\dot{G}_{N}/G_{N}|\leq6\times10^{-11}\
  yr^{-1}$, we find that the theoretical value $\dot{b}_{0}/b_{0}\simeq0.4$
  violates the upper bound $(\simeq0.14)$ by a factor $\lesssim3$.
\end{thebibliography}
\end{document}